\documentclass[pageno]{jpaper}

\usepackage{tikz}

\newcommand{\myparagraph}[1]{\noindent \textbf{#1}.}

\newcommand{\ourarch}{Violet}
\newcommand{\ourarchbase}{Vi2048}

\newcommand{\gminferlarge}{2.4X}
\newcommand{\powinferlarge}{10.6X}
\newcommand{\gmtrainlarge}{2.1X}
\newcommand{\powtrainlarge}{9.5X}
\newcounter{takeawaycount}
\newcommand{\tag}[1]{{\footnotesize \underline{$#1$}}}

\usepackage[normalem]{ulem}

\begin{document}

\title{Violet: Architecturally Exposed Orchestration, Movement, and Placement for Generalized Deep Learning}

\author{Michael Davies, Karthikeyan Sankaralingam\\
\{davies, karu\}@cs.wisc.edu}


\date{}
\maketitle

\thispagestyle{empty}

\begin{abstract}
Deep learning and hardware for it has garnered immense academic and industry interest in the past 5 years, with many novel proposals.
However, the state-of-art remains NVIDIA's TensorCore-based systems that provide top-of-line \textbf{performance} and \textbf{coverage} across a wide-spectrum of deep learning applications. 
In this paper, we first identify four key problems any new DL solution must solve: 1) Data orchestration, 2) Data movement, 3) Work placement and blending these to achieve 4) Coverage across different types of DL applications. 
With this as a guide, we propose \ourarch{}, a novel architecture with roots in multicore SIMD which balances the responsibilities for these four problems between the architecture, microarchitecture and software stack.
Compared to the NVIDIA A100 GPU, we find \ourarch{} achieves geo-mean \gminferlarge{}/\powinferlarge{} and \gmtrainlarge{}/\powtrainlarge{} performance/efficiency for inference and training across the MLPerf benchmark suite. 
We present detailed operator-level analysis of the MLPerf benchmark suite, extracting out key behaviors -- with implications for architecture research beyond this paper, that underpin the speedup and efficiency.
Overall, this paper motivates the importance of \textit{balance}, that the break down of responsibilities must be thought through carefully in order to compete with incumbent architecture designs.
\end{abstract}

\section{Introduction}
Deep Learning (DL) is one of the hottest topics in computing today, and its need for compute is insatiable. To meet this need, many styles of accelerator architecture are being explored, including NVIDIA's~\cite{noauthor_nvidia_nodate}, AMD's~\cite{noauthor_amd_nodate} and Biren's~\cite{biren-hotchips} GEMM engines, Google's TPU~\cite{jouppi2017datacenter}, dataflow and spatially programmed architectures like Xilinx Versal~\cite{noauthor_versal_nodate}, Graphcore~\cite{ltd_graphcore_nodate}, SambaNova~\cite{sambanova-mpr}, Groq~\cite{gwennap_groq_2020}, Qualcomm's AI-100~\cite{gwennap_qualcomm_2020}, and other proposals \cite{shao_simba_2019, chen2019eyeriss, venkatesan_magnet_2019, qin_sigma_2020}. There is also some delineation between training vs inference, and within that, support for particular types of DNNs (CNN, LSTM, GNNs etc.). 

Successful DL accelerators are quantified by their \textit{coverage} of DL applications, and \textit{performance, energy efficiency, and dollar cost} of those applications. Through this lens, NVIDIA’s TensorCore-based GPUs continue to be the dominant DL acceleration platform – supporting nearly every existing DL application as well as setting nearly every record for performance on the standard MLPerf~\cite{9138989} benchmark suite. Meanwhile, most other industry and academic contenders only report inference results for Resnet50 or BERT – and even in these cases, their performance is typically worse than NVIDIA. We term this apparent lack of performance and coverage from novel alternatives to GEMM acceleration, \textit{the DL accelerator gap}.

To help explain the mechanisms underpinning the DL accelerator gap, we identify, for any new DL accelerator architecture, four primary problems which must be addressed: 1) \textbf{work placement} – how work for a given DL operation is divided and assigned to the architectural resources, 2) \textbf{data orchestration} – how data required for some portion of work is moved to the compute resource, 3) \textbf{data movement} – how the compute resource handles unpacking and processing the data to carry out its task, and 4) \textbf{coverage} – how the architecture and its software stack support generality and scalability across the domain of DL. \textit{We take the view that \textbf{balancing} the responsibilities for these problems between hardware, programmer, and software stack is key to bridging the DL accelerator gap and achieve this balance by exposing it to the architecture. We revisit these four problems throughout the paper to show how they are interrelated and how our design solves them.}

\begin{figure}
    \centering
    \includegraphics[width=0.9\columnwidth]{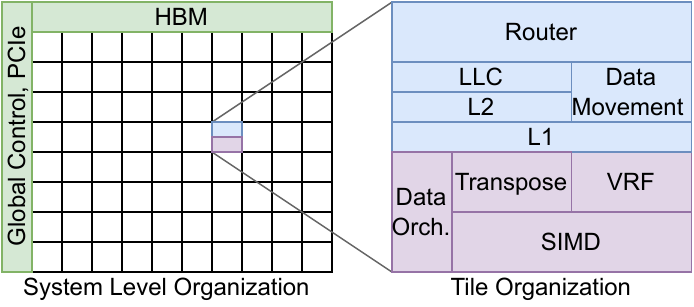}
    \caption{Overview of the \ourarch{} Architecture.}
    \label{fig:ourarch}
\end{figure}


Existing academic proposals for DL accelerators attempt to address these problems in a variety different ways but fall short. Data orchestration sees solutions ranging from relying on traditional SIMD processing \cite{venkatesan_magnet_2019, shao_simba_2019}, architecting PEs to be small and easy to keep active \cite{chen2019eyeriss} or complex software programmed networks at the core-level to deliver data to compute units \cite{qin_sigma_2020}. Data movement is typically solved by the interconnection on-chip -- some use a packet-switched NoC to make data movement entirely dynamic \cite{venkatesan_magnet_2019, qin_sigma_2020}, some introduce multicast primitives atop a traditional NoC \cite{shao_simba_2019} while others use an entirely statically configured NoC, moving the task of data movement entirely to software \cite{chen2019eyeriss}. In some cases,  relatively good balance of data orchestration and movement is achieved, for example, Simba~\cite{shao_simba_2019} and Sigma~\cite{qin_sigma_2020}. However, Simba is designed as an inference-only accelerator, and Sigma only evaluates GEMM operations. Table~\ref{tab:rel-work-summ} summarizes these works, as well as whether each of these four problems are solved by primarily software, hardware, or unaddressed.

\begin{table}[]
    \centering
    \caption{Summary of related work's solutions to DL acclerator challenges.}
    \label{tab:rel-work-summ}
\begin{tabular}{lcccc}
Architecture                          & \begin{tabular}[c]{@{}c@{}}Work \\ Place.\end{tabular} & \begin{tabular}[c]{@{}c@{}}Data \\ Mov.\end{tabular} & \begin{tabular}[c]{@{}c@{}}Data \\ Orch.\end{tabular} & Cov. \\ \hline
\multicolumn{1}{l|}{NVIDIA GPU}                                     & HW                   & HW                    & HW                                                    & Yes  \\
\multicolumn{1}{l|}{Simba (2019)~\cite{shao_simba_2019}}            & SW                   & HW                    & HW                                                    & No    \\
\multicolumn{1}{l|}{EyerissV2 (2019)~\cite{chen2019eyeriss}}        & SW                   & SW                    & HW                                                    & No    \\
\multicolumn{1}{l|}{MAGNet (2019)~\cite{venkatesan_magnet_2019}}    & SW                   & HW                    & HW                                                    & No    \\
\multicolumn{1}{l|}{SIGMA (2020)~\cite{qin_sigma_2020}}             & SW                   & HW                    & SW                                                    & No    \\
\multicolumn{1}{l|}{\ourarch{}}                                      & SW                   & HW                    & HW                                                    & Yes   \\ \hline
\end{tabular}
\end{table}

In this work, we develop a novel accelerator architecture, \ourarch{}, with DL coverage as a first-order goal, achieving this through carefully balancing the responsibilities for work placement, data orchestration, data movement between the architecture, microarchitecture and compiler. Fine grained data orchestration is supported by extending traditional SIMD with an architecturally exposed transpose engine, enabling highly efficient execution of state-of-art DL workloads by reusing cache lines loaded from a core's private data cache. Efficient data movement is tackled by extending a 2D mesh NoC to be a first-class programmable component in the architecture with the inclusion of a special data movement core. Flexible work placement is supported by the kernel-based execution model coupled with a dynamic, cache-based, memory system. Finally, coverage of DL applications is easily achieved as a result of our choice of balance lending to rapid software stack development -- and this coverage is demonstrated by \ourarch{}'s ability to run all of the applications (inference and training) in the MLPerf benchmark suite. 

Specifically, the contributions of this work are:

\begin{itemize}
    \item A detailed qualitative and quantitative characterization of a broad set of DL applications and their imputed needs on the hardware/architecture. We evaluate 300+ different shapes of operators, across CNNs, Transformers, and LSTMs which as far as we know is the widest such study.
    \item A novel architecture, \ourarch{}, designed to balance the responsibilities of work placement, data orchestration, and data movement between the architecture and software-stack. We find that for data orchestration in particular, a narrow subset of AVX plus a small extension suffices.
    \item A detailed distillation of how \ourarch{}’s architectural features and balance of responsibilities enable \ourarch{} to be an easy compilation target across a broad set of DL workloads in addition to achieving competitive or superior performance compared to state-of-art industry solutions. We call out specifically what features of each operator can be leveraged by \ourarch{}'s exposed data orchestration and movement primitives.
    \item An analysis across DL applications, comparing \ourarch{} to the NVIDIA A100 GPU which shows \ourarch{} achieves \gminferlarge{} / \gmtrainlarge{} geo-mean speed-up at batch 16 size inference / training with \powinferlarge{} / \powtrainlarge{} power efficiency.
    \item A deep dive into operator-shape level characteristics and behaviors across MLPerf. In particular, we distill out 8 key behaviors which have generalizeable implications beyond just our proposed design, and could even be used to help improve GPUs. Table~\ref{tab:key-behav-summ} summarizes the key behaviors.
\end{itemize}

\begin{table}[]
    \centering
    \caption{Summary of key behaviors in DL operator shapes.}
    \label{tab:key-behav-summ}
\begin{tabular}{ll}
\#                     & Behavior                                                                                          \\ \hline
\multicolumn{1}{l|}{1} & Large-channel convs. that map well to GEMM units                                                 \\
\multicolumn{1}{l|}{2} & Large-spatial convs. that give easy parallelism                                    \\
\multicolumn{1}{l|}{3} & Unit-filter convs. which are just matrix multiply                               \\
\multicolumn{1}{l|}{4} & Conv. backpropagation which is hard to parallelize                                              \\
\multicolumn{1}{l|}{5} & Choice of tiling can impact comm. via placement                                             \\
\multicolumn{1}{l|}{6} & Large matmuls. that are easy to get high perf.                                \\
\multicolumn{1}{l|}{7} & Odd-shaped Batch matmuls. hard data orch. \\
\multicolumn{1}{l|}{8} & LSTM with low parallelism                \\ \hline
\end{tabular}
\end{table}

The rest of this paper is organized as follows.
Section~\ref{sec:rel-work} explains further the related works to this paper and our differentiation from contemporary DL accelerator designs.
Section~\ref{sec:background} overviews and analyzes DL applications, providing a broad set of workload behaviors and how they frame the four key problems for bridging the gap.
Section~\ref{sec:arch} presents \ourarch{}, a novel architecture for DL acceleration that achieves high performance and coverage of state-of-art DL apps in the MLPerf benchmark suite.
Section~\ref{sec:mapping} shows how \ourarch{}'s balance of responsibilities and solutions to the four key problems make it possible to rapidly produce performant mappings of dominant DL operators.
Section~\ref{sec:methodology} describes our methodology for evaluating \ourarch{}.
Section~\ref{sec:results} presents our evaluation, where we explore the design space of \ourarch{}, its performance and efficiency compared to the A100, the key application behaviors of MLPerf operators that afford this performance, as well as a limit study on the possible future improvements for \ourarch{}.


\section{Related Work}\label{sec:rel-work}
\myparagraph{\ourarch{} Positioning} Within the space of platforms for DL, general purpose processors (GPP) are one end, achieving low performance, high coverage and easy compilability; GPUs are in the middle achieving high performance, efficiency, coverage and good compilability by adding specialized units to an existing flexible architecture; DL accelerators use exotic architecture, aspiring for extreme performance efficiency, and have thus far sacrificed generality and make compilability hard~\cite{shao_simba_2019, chen2019eyeriss, qin_sigma_2020, venkatesan_magnet_2019}. Table~\ref{tab:qual-comp} summarizes these observations. As argued in~\cite{sankaralingam_mozart_2022, gwennap_groq_2020}, compilability -- the ease in which DL operations can be lowered to an architecture -- is a fundemental requirement to usability.
This work explores the GPP paradigm to answer whether we can get higher efficiency than a GPU while also providing compilability. REDUCT~\cite{nori_reduct_2021} is the closest \textit{philisophically} related work to \ourarch{}. 

\begin{table}[]
    \centering
    \caption{Qualitative comparison of General Purpose processor, GPU and AI Accelerators.}
    \label{tab:qual-comp}
\begin{tabular}{llll}
\hline
                                    & GP Core & GPU        & AI Acc.   \\ \hline
\multicolumn{1}{l|}{Efficiency}     & Low     & High       & Higher      \\
\multicolumn{1}{l|}{DL Generality}  & High    & High       & Low       \\
\multicolumn{1}{l|}{HPC Generality} & High    & High       & Very Low  \\
\multicolumn{1}{l|}{Compilability}        & Easy    & Autotuning & Hard      \\ \hline
\end{tabular}
\end{table}

\myparagraph{DL Accelerators} There are many proposed designs for DL acceleration \cite{chen2019eyeriss, qin_sigma_2020, shao_simba_2019, venkatesan_magnet_2019}. These architectures are all based on an array of PEs which contain structures optimized for multiply-accumulate (MAC) or GEMM operations (or SpMM to exploit structured sparsity). 
EyerissV2~\cite{chen2019eyeriss} uses a circuit switched, statically configured NoC, meaning software must plan all data movement at compile time. Further, the way in which this NoC is exposed architecturally is not examined and only configurations for convolution and matrix multiply inference are evaluated. 
Sigma~\cite{qin_sigma_2020} and MAGNet~\cite{venkatesan_magnet_2019} both have similar system level architecture. They use a traditional 2D mesh, relying on packet-switched routing for data movement and carefully crafted work placement to optimize communication pressure. Sigma uses a custom PE design with software configured data orchestration. MAGNet uses a conventional SIMD-style compute unit. Sigma only evaluates matrix multiply workloads, constraining it to DNNs such as GNMT and Transformers. MAGNet only considers convolution inference, and is intentionally specialized this way.
Simba~\cite{shao_simba_2019} is designed to be a small-batch inference accelerator based on chiplets, with a mesh NoC on each chiplet, and mesh NoC on the whole package. Simba also has a ``Global PE'' for near-data operations but only this one compute component is able to perform this kind of work. 
Table~\ref{tab:related-work} summarizes the differences between these works and this paper. 


\begin{table}[]
    \centering
    \caption{Summary of \ourarch{}'s differences to related work.}
    \label{tab:related-work}
\begin{tabular}{lll}
\cline{1-2}
\multicolumn{1}{c}{Architecture}      & Comments or differences to our approach                                        &  \\ \cline{1-2}
\multicolumn{1}{l|}{Simba}     & Chiplets, Special bufs, Inf. for CNN only &  \\
\multicolumn{1}{l|}{EyerissV2} & Special PE, bufs, HM-NoC, \\  \multicolumn{1}{l|}{}  & Inf. for CNN only  &  \\
\multicolumn{1}{l|}{MAGNet}    & RTL Gen., Special bufs, Inf. for CNN only   &  \\
\multicolumn{1}{l|}{SIGMA}     & Specialized PE, DNN only              & \\ \cline{1-2}
\end{tabular}
\end{table}

\myparagraph{SIMD Architectures} There have been recent works on SIMD and in particular, looking at AVX extensions. These include REDUCT~\cite{nori_reduct_2021}, analysis of convolution performance~\cite{georganas_anatomy_2018}, and analysis of inference on CPUs~\cite{liu_optimizing_2019}. Mittal et al.~\cite{mittal_survey_2021-1} presents a survey of deep-learning on CPUs and focus on issues of memory hierarchy and datapath. Domke et al.~\cite{domke_matrix_2021} present a thoughtful case to understand and revisit the role of Matrix Engines for HPC. Reuther et al. present a survey on DL accelerators~\cite{reuther_survey_2020-1}. These works do not look at the details of program behavior, contribution of architecture, or DL coverage. Google has published an extensive set of papers on TPU including~\cite{jouppi_ten_2021}, which cover the software stack, systolic array architecture, and datatypes. 

\myparagraph{DL Application Analysis and Software Techniques} Verma et al. present a workload characterization of MLPerf Training~\cite{verma2019demystifying}. Cross-layer approaches related to our work include high- \cite{astra} and low- \cite{in-register, noauthor_amos_2022} level code generation techniques, and also memory management \cite{swap-advisor} and memory partitioning techniques \cite{mem-centric, accpar, split-cnn}. Operator mappers which design dataflows for operator shapes employ techniques such as loop ordering and search~\cite{parashar_timeloop_2019, kwon_maestro_2020, huang_cosa_2021}, intrinsic mapping~\cite{noauthor_amos_2022}, template-based~\cite{georganas_anatomy_2018, noauthor_cutlass_nodate}, and manual programming~\cite{chetlur_cudnn_2014, eigenweb}.

\section{Characterization of DL Applications}\label{sec:background}

\begin{figure}
    \centering
    \includegraphics[width=0.9\columnwidth]{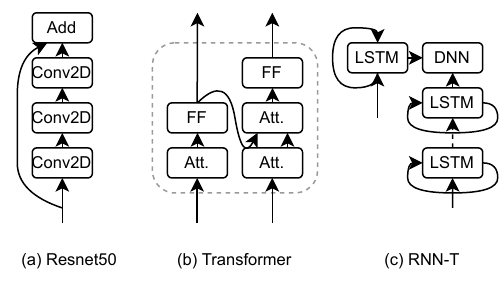}
    \caption{Overview of Common DL Applications}
    \label{fig:workload-graphic}
\end{figure}

This section provides an overview of deep-learning applications focusing on distilling the program behaviors and the implications on the four DL accelerator challenge problems. 
\subsection{An Overview of a DL Application}
Deep learning applications are computer programs which operate over \textit{tensors} -- multidimensional arrays. Tensors are  either \textit{fixed} or \textit{learned}. Fixed tensors are given (unchangeable) inputs to a network, and learned tensors are called \textit{parameters} or \textit{weights}. During the execution, a directed acyclic graph (DAG) is constructed where nodes are operations, and edges are the tensors that are consumed or produced by that operation. This \textit{compute graph} represents the dynamic instance of computation performed by the application, and serves as the input to the automatic differentiation (autograd) algorithm to compute compute gradients of every tensor with respect to a given output (the ``loss'' value). The initial execution of the program and construction of the compute graph is referred to as \textit{inference} or the \textit{forward pass}. Autograd, and gradient computation is often referred to as \textit{backpropagation} or the \textit{backward pass}. Both forward and backward passes together are referred to as \textit{training}. DL applications produce the same compute graph for any input. These applications can be captured by its compute graph alone\footnote{In some well behaved, non-static cases, such as recurrent networks, we can still capture the compute graph by introducing cyclic edges. These are called ``quasi-static''.}. Figure~\ref{fig:workload-graphic} shows the compute graph for several example DL applications. Most DL applications are implemented using frameworks such as Tensorflow, PyTorch and ONNX. These frameworks, in addition to providing implementations for operators, handle automatic differentiation. To support new architectures, the onus is on the accelerator designer to lower framework-level operators to their architecture, creating a substantial software lift necessitating compilability as a first order requirement

\subsection{Characterization and Implications for Hardware}
From a computer architecture perspective, the operations' semantics, their execution order, as well as tensor \textbf{shape} (dimensions), \textbf{layout} (the order in which these dimensions are flattened in memory) and \textbf{datatype} play a role in data orchestration, data movement and work placement. Here we characterize the entire MLPerf suite and detail how these characteristics impact these problems. Table~\ref{tab:workload-properties} summarizes the quantitative features of each of the applications. Based on qualitative understanding of the applications and detailed quantitative profiling (methodology explained in Section~\ref{sec:methodology}), our general findings are below.

\begin{table}[]
    \centering
    \caption{Summary of different properties of workloads}
    \label{tab:workload-properties}
    
\begin{tabular}{lrrllr}
\hline
\multicolumn{1}{c}{Network} &
\multicolumn{1}{c}{GOPs} &
\multicolumn{1}{c}{Shapes} &
\multicolumn{1}{c}{Primary Ops} &
\multicolumn{1}{c}{\%} \\ \hline

\multicolumn{1}{l|}{RN50} & 8 & 23 & Conv2D & 99\% \\
\multicolumn{1}{l|}{SSD}  & 427 & 30 & Conv2D & 99\% \\
\multicolumn{1}{l|}{UNET} & 938 & 18 & Conv3D & 99\% \\
\multicolumn{1}{l|}{BERT} & 110 & 5 & MatMul & 98\% \\
\multicolumn{1}{l|}{RNNT} & 14 & 6 & LSTM, MM & 94\% \\
\hline
\multicolumn{1}{l|}{RN50} & 24 & 69 & Conv2D & 99\% \\
\multicolumn{1}{l|}{SSD}  & 83 & 90 & Conv2D & 99\% \\
\multicolumn{1}{l|}{UNET} & 2816 & 54 & Conv3D & 99\% \\
\multicolumn{1}{l|}{BERT} & 479 & 15 & MatMul & 98\% \\
\multicolumn{1}{l|}{RNNT} & 42 & 14 & LSTM, MM & 94\% \\
\hline
\end{tabular}

\end{table}

\myparagraph{Operators and Application Coverage} Across the MLPerf suite of applications, three operators dominate: matrix multiply, convolution, and LSTM, accounting for over 90\% of all ops in each network. For these three, over 300 unique shapes exist with various amounts of arithmetic intensity, available reuse, layouts, intermingling with elementwise operations such as ReLU, batchnorm all with different variations for forward and backward pass. This means any solution for coverage directly depends on its solutions for work placement, data orchestration and data movement across the set shapes to be supported. In addition, while focus can be placed on these operations, an architecture needs to be balanced to support the range of DL operations needed (E.g. Batch Norm, Layer Norm, Softmax) otherwise it will be limited by Amdahl's law at best, or be unable to achieve DL application coverage at worst. \textit{A good solution to DL coverage is a composite of solutions to data orchestration, data movement and work placement, and how these solutions generalize.}

\myparagraph{Layout and Datatype drives Data Orchestration} Tensor layout directly impacts which dimensions can be used for vectorization and what minimum tiling factors are needed. For some datatypes (Integer as well as Float16), multiply-accumulates use a wider datatype for accumulation. For example, Intel's recent AVX extensions support an Int8 to Int32 multiply-accumulate~\cite{cueva_code_2019}. On the hardware side Int8 to FP16 costs 3X in area, up to 5X in power, while Int8 to FP32 costs 10X to 20X in area and power~\cite{abdel-aziz_rethinking_2021,johnson_rethinking_2018,zhang_flexible_2019,fleischer_scalable_2018}. 
\textit{A good solution to data orchestration must be aware of, and be performance-agnostic to layout or datatype, and in the process, avoid introducing needless software or hardware complexity. Care must also be taken to balance data orchestration tiling needs with work placement parallelization needs (explained below) to achieve maximum utilization.}

\myparagraph{Reuse drives Data Movement} Arithmetically intense DL operations (all three of our dominant operators typically have high arithmetic intensity) afford a lot of data-reuse opportunities -- and for most DL operations, data-dependencies are entirely statically defined. In addition, having a static (or quasi-static) compute graph allows operations to be executed in topological order. This means that often the output from one operation will be immediately used in the next operation lending to an obvious temporal locality of tensor operands. 
\textit{A good solution to data movement must be able to recognize and exploit available information on data dependencies and reuse. While it may seem intuitive to solve entirely with software, this approach is typically involves leaking microarchitectural constraints to the software creating performance pitfalls when trying to generalize. Balance must be struck between data movement and application coverage.}

\myparagraph{Parallelism drives Work Placement} The amount of and ease in exploiting available parallelism are the key factors which impact work placement. The three dominant operators we observe all have ample parallelism. In addition, for inference, batching of multiple inputs provides higher reuse and more embarrassing parallelism, with almost no additional pressure on the hardware. \textbf{Training with large batches, provides the same, but a linear increase in the amount of intermediates that need to be kept-around, before the backward pass can commence, meaning larger memory capacity is needed resulting in higher total dollar cost as well as higher memory power.} Work placement also directly impacts data movement -- units of work which share input data are best placed physically proximate to eachother to ease the burden on data movement and allow it to extract out broadcasting opportunities. \textit{A good solution for work placement should support coverage by abstracting microarchitectural constraints, strategically exposing constraints when critical for data movement. It should also afford tuning placements for extracting additional performance with expert knowledge.}



    

    


\section{\ourarch{} Architecture}\label{sec:arch}
In this section we detail the design of \ourarch{}, a novel architecture with roots in multi-core SIMD, for generalized deep learning acceleration. We first overview the organization of \ourarch{}, then explain in detail how \ourarch{}'s design solves the four problems of data orchestration, data movement, work placement and application coverage.

\begin{figure}
    \centering
    \includegraphics[width=1.0\columnwidth]{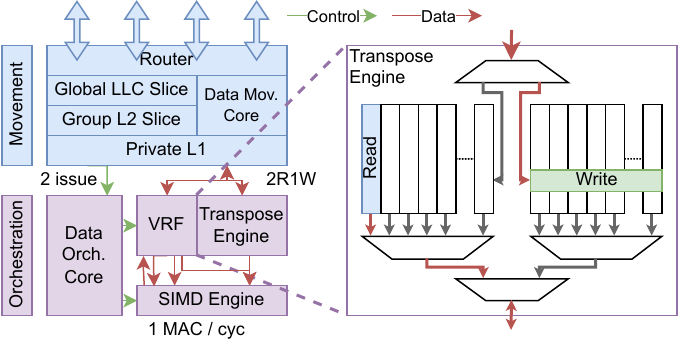}
    \caption{Organization of \ourarch{} Tile.}
    \label{fig:arch-detailed}
\end{figure}

\subsection{System Organization} 
\ourarch{}, shown in detail in Figure~\ref{fig:ourarch} and Figure~\ref{fig:arch-detailed}, consists logically of three main components: 1) Parallel processing elements 2) An interconnection network and 3) A memory system. Physically, \ourarch{} is divided into identical tiles, each containing a data orchestration core coupled with a wide SIMD/short-vector datapath including register file and arithmetic units organized as lanes\footnote{From here on, we use SIMD, SIMD Register-File and SIMD lane}. The tile also contains a slice of a distributed memory hierarchy over a 2D mesh NoC combined with a data movement engine. We find the mechanics of the ISA are unimportant as suggested by Blem et al.~\cite{isawars}.

The system includes a global thread scheduler that transmits work to cores based on software-defined work placement. It also includes a host interface controller (PCIe-like interface) to provide high bandwidth, low latency communication to a general purpose host computer that runs the system-level portions of the DL stack. Finally, one or more memory controllers and PHY on chip (HBM from a implementation standpoint is preferred) feed the LLC. The physical organization of the LLC is straight-forward: slices distributed across the chip with static address mapping\footnote{Tensor layout can be optimized to fine tune proximity of slices to cores}. A 2D-mesh interconnection network transmits cache lines between tiles and to and from memory. 

\subsection{Data Orchestration}\label{sec:arch-orch}
Data orchestration concerns how the compute resources of an accelerator consume data and map it to execution resources. Quantitatively, data orchestration is solved \textit{well} if compute resource utilization under \textit{ideal memory} conditions is high. \ourarch{}'s SIMD datapath supports a small set of conventional SIMD instructions: add, multiply, multiply-accumulate, vector load/store (with broadcast \& stride), wide-accumulate. In addition, we introduce a transpose engine, a novel microarchitecture component, which exposes custom vector instructions for loading transposed 2D blocks of vector elements while maintaining memory throughput. Section~\ref{sec:mm-deepdive} further explains the use and benefit of the transpose engine. The transpose engine relies on minimum block size to amortize cache-line loads over the number of resulting SIMD vectors it produces. We synthesized the transpose engine in RTL to confirm the designs feasibility. One SIMD lane supports two Int8 and one FP16 multiply-accumulate operation per cycle. It allows internal accumulation in 32-bits. The size of the architectural register file and SIMD lanes is a first-order determinant of performance, since it dictates how much parallel work can be done before hitting WAW hazards. We found that with 32 registers, MLPerf applications can be supported without the core becoming the bottleneck.

\subsection{Data Movement} 
Data movement concerns how data needed for computation is moved from storage (typically, external DRAM) into the compute resources. Quantitatively, data movement is solved \textit{well} if end-to-end compute resource utilization under \textit{ideal compute} conditions is high. \ourarch{} addresses data movement by combining a traditional memory hierarchy with a special communication interconnect that includes a programmable data movement engine. This programmable network allows software to facilitate a ``push''-style of prefetch operation, where the LLC essentially can ``push'' data to destinations over the network automatically (and without software planned routes), eliminating request traffic. The rich information in DL stacks allow such static analysis to be effective and straightforward (unlike codebases like SpecINT, SpecFP etc.). The three-level hierarchy of our memory system and L2 group sizes are chosen to enable the data movement engine to specify all destinations in the packet header, allowing the NoC and routing algorithm to intelligently multicast cachelines at any router, reducing network traffic to transmit a cacheline. In addition to this, \ourarch{} is able to support work that operates directly on the local LLC slice in a tile, meaning element-wise operations do not require data movement at all. The tile's private cache is sized to be 32 KB as a staging area for data. 



\begin{figure}
    \centering
    \includegraphics[width=0.8\columnwidth]{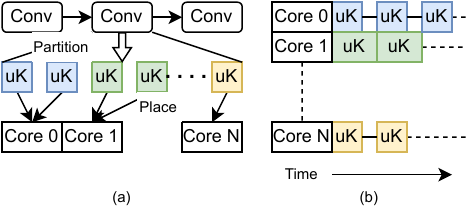}
    \caption{Execution model and lower of graph to hardware primitives. (a) Shows how DL operators are lowered down to low-level primitives. (b) The execution model of an operator. }
    \label{fig:execution-model}
\end{figure}

\subsection{Work Placement}
Work placement describes how an operation's work is allocated to available resources on the architecture. A placement solution must often be spatially aware to understand what placement options are best for spatial data locality (and thereby improve data movement efficiency). It must also be aware of the details of the memory system (coherence, consistency, etc) as well as the execution model, to understand what work is allowed to be placed on what resources. For the purposes of work placement, \ourarch{} can be viewed as a parallel thread array with one thread per core, with incoherent memory between physical cores. To lower an operator to \ourarch{}, a programmer or software \textit{partitions} work for a given operator into individual tasks, each of which run on one logical thread. They then \textit{place} this work onto the available cores with goal to map tasks with overlapping sets of data onto the same core (to ease data movement) while also balancing parallelism. Figure~\ref{fig:execution-model} provides a pictorial representation of this process.

\section{Demonstrating Coverage}\label{sec:mapping}

\begin{figure*}
    \centering
    \includegraphics[width=0.8\textwidth]{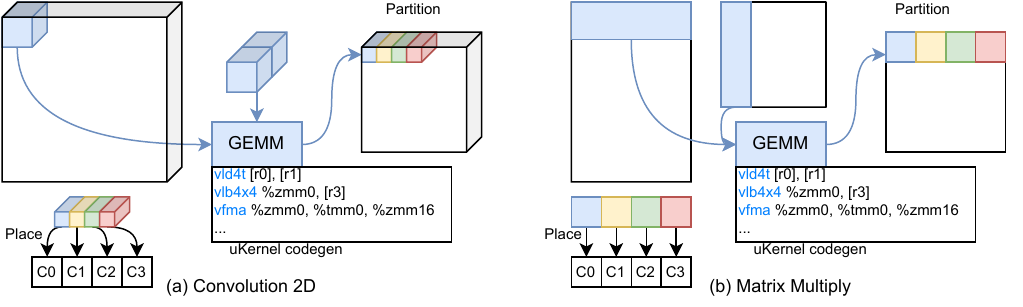}
    \caption{Mapping of Common Operators to the \ourarch{} architecture}
    \label{fig:algo-mapping}
\end{figure*}

For \ourarch{} to achieve coverage of DL applications and demonstrate the compilability of \ourarch{}, we developed techniques for mapping and lowering the primary operators from MLPerf, paying close attention to the role of \ourarch{}'s novel transpose engine and programmable interconnect in achieving high performance and developer user experience. 

We adopt output-stationary dataflow as a baseline strategy. This lends to an easy to understand parallel algorithm in that it eliminates inter-thread communication, leaning on ``push'' prefetch and dynamic multicast to exploit load-reuse opportunities, and the transpose engine to deliver performant data orchestration. Work items are sized to optimize for arithmetic intensity, respecting minimum tiling parameters to fill the SIMD+Transpose unit, as well as L1 data cache size. For each operator, we develop a micro-kernel which handles a single output chunk. Figure~\ref{fig:algo-mapping} shows an overview of operator mapping for two representative operators: convolution and matrix multiply. It shows the highest level operations in terms of the two tensors, the chunking to achieve parallelism, reuse available, and the lowest-level code snippet. We now explain each operator's mapping in more detail, calling out the key details which \ourarch{} leverages to achieve high performance and efficiency without a complex software lift.

\myparagraph{Matrix Multiply}\label{sec:mm-deepdive}
\begin{figure}
    \centering
    \includegraphics[width=0.8\columnwidth]{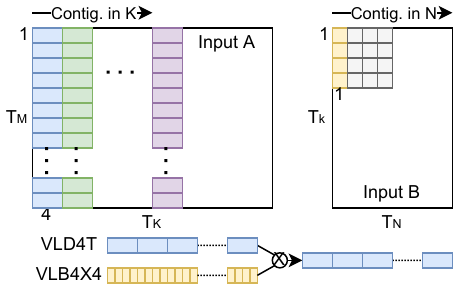}
    \caption{Use of transposed-loads in a small matrix-multiply.}
    \label{fig:matmul-transposed}
\end{figure}
Linear layers in DL applications are implemented as a matrix multiply of two input matrices $A$ and $B$ which have shapes $A[M, K]$ and $B[K, N]$ (B is typically the ``weight'' tensor). We base our strategy on NGemm~\cite{bao2019ngemm}, taking into account the effects of wide-accumulation for integer datatypes. Each output chunk can itself be computed as a matrix multiply of slices of the original $A$ and $B$, and the minimum size of these chunks is dictated by which dimensions are vectorized over and by how much. \ourarch{}'s Transpose Engine allows for a special type of ``transposed-load'' which allows a number of cache lines to be loaded with their data then striped across many transposed vector-registers. The number of cache lines loaded is equal to the vector width (in bytes) divided by the ratio between the input and accumulation datatypes ($4$ in the case of Int8->Int32 matrix multiply). \textbf{This special transposed-load is employed to reduce the number of cache line loads needed to fill vector registers with relevant data by exploiting spatial locality in the algorithm. The exact layout combination -- that is, whether A, B, both, or neither are transposed themselves -- impacts how transposed-loads are employed and ultimately, the minimum tiling factors needed for sustained throughput.} Figure~\ref{fig:matmul-transposed} depicts the use of a transposed load (VLD4T) and a broadcast load (VLB4X4, similar to vbroadcastss in AVX-2) to fill vector registers to be used in a multiply-accumulate operation.

From another perspective, using the transpose engine changes the semantics of the SIMD MAC from VL-dot products of size 1 to (VL/R)-dot products of size R (R=4 for Int8, R=2 for Float16). This allows a tradeoff between the minimum K needed and minimum N needed in order to hide memory loads behind vector MAC operations. 
In the case of MKKN layout in Figure~\ref{fig:matmul-transposed}, if we have $T_M=16$, $T_N=4$, $T_K=16$, we are loading 16 cache lines from A, 4 cache lines from B, and performing 16 vector MACs. With \ourarch{}'s dual read-port cache, we can cover the 10 cycles needed to load cachlines behind the 16 vector MAC operations. The work placement algorithm we employ for \ourarch{} understands this tradeoff and is able to select which minimum tile is better for a given shape.

Data movement of input slices to a core is part of the microkernel specification. A mechanical process is employed to enumerate the slices needed by each core across the timesteps of execution and generate a data movement program which pushes relevant data in the local LLC and L2 slice to each of the consumers that need it.

Back-propagation is also a matrix multiply operation, multiplying the output gradient by the original two inputs in two separate operations -- and in each case, the layout of one of the tensors is transposed. So for a forward pass that is MKKN, the backward pass will observe layouts of KMKN and MKNK. We similarly can choose min. tiling factors based on what is needed to cover cache line loads with vector MAC operations.

\myparagraph{Convolution}
We implement convolution based on Intel's approach~\cite{georganas_anatomy_2018}. We similarly adopt an output-stationary dataflow. Each output block is computed by invoking a small GEMM kernel over the input and filter, with the reduction dimension being the input channels of the convolution. \textbf{Because the microkernels for convolution are small matrix multiplies, we can reuse the same analysis for matmul to define minimum tile sizes for maximum compute throughput.} 

For computing convolution input back-propagation (dI) and weight back-propagation (dW) we follow a similar approach to designing an algorithm -- employing our small matrix multiply algorithm to compute output chunks for these two operations. Weight gradient computation has to reduce over spatial dimensions, so there is little available parallelism. We adopt a similar strategy to \cite{georganas_anatomy_2018} in this case, applying some tiling factor to the batch dimension and computing partial gradients over this tiling factor. We then apply a reduction kernel to sum the partial gradients. Using \ourarch{}'s ability for core to operate directly out of the LLC, we are able to perform this reduction without incurring any additional communication cost.

\myparagraph{LSTM}
\textbf{The LSTM blocks in RNN-T decompose into two very nice operations: a matrix multiply, and an elementwise operation.} The matrix multiply is for the linear layer that transforms the input and hidden state for the current time step of the LSTM. The LSTM operation, past this linear layer, is elementwise over the vector elements of the input and hidden state. \textbf{We employ our high performance matrix multiply algorithm to handle the linear layers, then use the same in-LLC compute to handle the LSTM Cell's element-wise computation.} Backpropagation is quite simple in that it just requires an element-wise gradient operation for the LSTM computation, and then a matrix-multiply backpropagation which we can reuse the algorithm from before.


\subsection{Compiler and Software Stack}
\ourarch{}'s software stack is positioned to support a DL-framework level abstraction. At the top level is TensorFlow and PyTorch which issue calls to allocate and transfer memory, and execute operators. Our software stack is made up of a runtime component which handles memory management and device excution, and a library component which contains operator code templates that are specialized just-in-time based on operator shapes. These templates are stored in a domain specific language that separates out the mapping task into low-level code-generation and work placement. Code-generation is handled by a conventional compiler, while work placement is handled by the above strategies. The architecturally exposed mechanics for data orchestration and movement can be lowered to easily by this template-based approach. Thus, we address the compilability challenge that plagues other designs.

\section{Methodology}\label{sec:methodology}
We now detail our evaluation methodology of end-to-end
DL applications implemented in TensorFlow/PyTorch.


\myparagraph{Performance Modeling}
We insert region markers into TensorFlow and PyTorch at the operator boundary to generate an operator trace. For performance evaluation, we build a Zsim-like performance simulator/model that uses the DL operator trace, and a memory system model, accounting for the effects of the vector instructions, access rates to private memory, as well as NoC contention.

\myparagraph{Power and Area Estimation}
We used the methodology in Accelergy~\cite{wu_accelergy_2019} and Timeloop for our area and power modeling. We use the LX3 processor core mentioned in \cite{nowatzki2016pushing,halfhill_xtensa_2009} as a reference for the power and area of a lightweight, in-order core. Ara \cite{cavalcante_ara_2019} provides an estimate of SIMD area and power. We use the arithmetic units from \cite{johnson_rethinking_2018} as a reference for the SIMD MAC unit. Finally, Cacti is used for power and area estimates of the last level cache. All of these components are normalized to 7nm power and area using the methods from \cite{stillmaker_scaling_2017}. The power consumption of the memory controllers, PHY and HBM stacks is 6 Watts per stack (24W for entire chip) based on data sheets for HBM2. For frequency scaling, we make use of work presented by ARM on their Neoverse N1 CPU, which presents power scaling for 3 GHz to 1.2 GHz \cite{9062889}.


\myparagraph{Baseline and Comparing Performance} To report our results, we obtained published performance results of the NVIDIA A100 system. We paid close attention to ensure that we were using the exact same DL-model as the NVIDIA system. For some applications, we used NVIDIA's code published through MLPerf to replicate performance results and obtain results for different batch sizes. For Bert Large pretraining, we were unable to run NVIDIA's official code on a single GPU so we used the ratio of large batch inference to small batch inference to estimate small batch pretraining. To obtain layer-level performance and efficiency, we ran individual operators with PyTorch on an A100 and collected runtime using NVIDIA NSYS, computing utilization with FLOP counts for each layer.


\myparagraph{Limitations}
We focus on an execution model of one operator active at a time -- leaving inter-operator parallelism for future work. As described in the results, this simpler approach provides substantial performance and efficiency already. Also, we focus on single-node training, with the observation that techniques for high-performance distributed training are orthogonal to single-node performance. Klenk et al. show that perfect all-reduce improves performance by 10\% to 40\% ~\cite{klenk_innetwork_2020}. Finally,
we present qualitative comparison to existing academic designs since they don't support many of the operators in MLPerf for full application execution and comparison, precluding a ``fair'' quantitative comparison against them.


\section{Results}\label{sec:results}
We evaluate \ourarch{} across the MLPerf benchmark suite, gathering detailed performance and power data at an operator level granularity. The results of our study are organized as follows.
Sec~\ref{sec:dse} introduces \ourarch{}'s design space, examining what design parameters have the biggest impact on overall performance. 
Sec~\ref{sec:comparison-a100} presents a comparison of our chosen configuration, \ourarchbase{}, to an NVIDIA A100 GPU -- a state-of-art DL accelerator -- showing how \ourarchbase{} can achieve superior performance and power efficiency. 
Sec~\ref{sec:shape-analysis} dives deeper into the operator shapes which have the highest impact on run time for each workload, distilling how the balance of responsibilities for data orchestration, data movement and work placement help enable high performance on each shape.
Sec~\ref{sec:sens-scale} presents a scalability study, examining the effects of optimizing different components of \ourarch{}, for the purpose of elucidating where the bottlenecks are, and what components should be focused on for further improvement.
\textbf{We evaluate the entire 300+ unique operator shapes in the MLPerf application suite, extracting out performance and efficiency characteristics.}

\subsection{Design Space Exploration}\label{sec:dse}

\begin{figure*}
    \centering
    \includegraphics[width=1.0\textwidth]{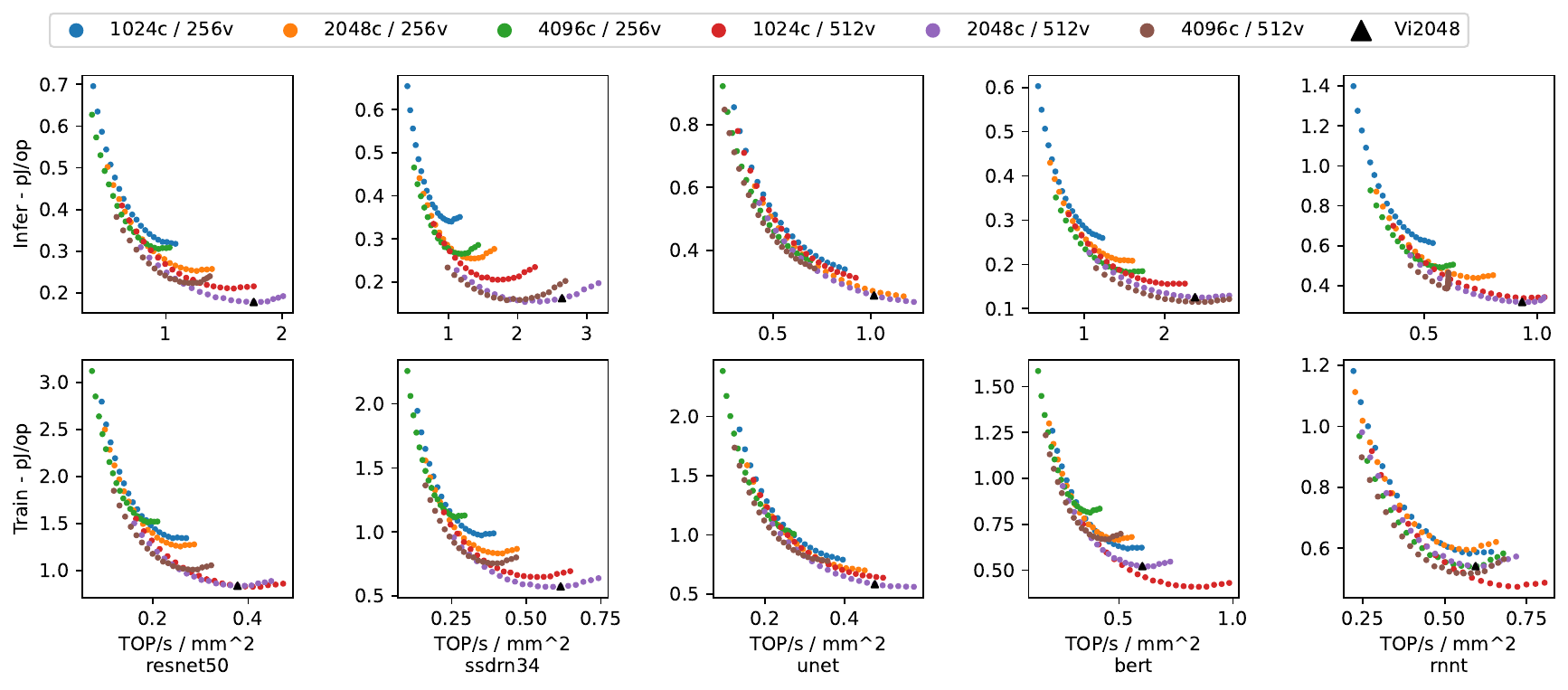}
    \caption{Design Space Exploration of \ourarch{}. Points of same color sweep from 1GHz-3GHz clock frequency with steps of 100MHz.}
    \label{fig:dse}
\end{figure*}

We first conducted an in-depth analysis of the design space that is created over the parameters of SIMD-width, number of processing cores, and frequency. We used our model to plot all of the design points in our space in terms of their area efficiency (TOPS / $mm^2$) and power efficiency (pJ / OP). Figure~\ref{fig:dse} shows the results of this survey. We can see that there is quite a diverse spread of design points that vary by up to a factor of 3X in terms of area efficiency, and 4X in terms of power efficiency. Further, we observe that not all applications agree on which design point is the most efficient overall. 

To elucidate efficiency, we search design space and consider only points that match peak performance of A100. This allows us to look at utilization, speedup and perf/watt as metrics to evaluate underlying the efficiency of the architecture. For each application, we then identified the best configuration in terms of power efficiency, and then in terms of area efficiency to break ties. The result was \ourarchbase{} -- 2048 cores, 512-bit SIMD, at 2.4GHz operating frequency. We also observe that \ourarchbase{} comes within 20\% of the optimal TOPS/$mm^2$ and within 25\% of the optimal pJ/OP for each applications, with the exceptions of 50\% for BERT and RNN-T large batch training. 


\subsection{Comparison to A100}\label{sec:comparison-a100}


\begin{figure}
    \centering
    \includegraphics[width=1.0\columnwidth]{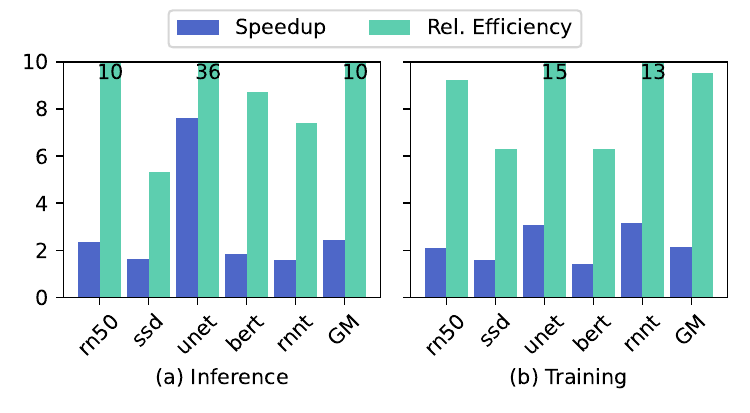}
    \caption{\ourarch{} vs A100 across MLPerf at batch size = 16. Throughput in samp/sec, and Rel. Eff. is ratio of avg. pJ/op.}
    \label{fig:perf-compare}
\end{figure}


\begin{table}
    \centering
    \caption{Comparison of Specs.}
    \label{tab:spec-comparison}
    \begin{tabular}{c|c|c}\hline
    Spec & \ourarchbase{} & A100 \\ \hline
    Die Area     & 215 $mm^2$ & 840 $mm^2$ \\
    Peak TOPs (Int8/FP16)     &  524/262 & 624/312 \\
    Power (TDP) & 100 W & 300 W \\
    Frequency & 2.4 GHz & 1.4 GHz \\ 
    Total L1/L2/LLC Size & 64/32/128 MB & 20.25/-/40 MB \\
    HBM2 Memory & 16 GB & 40 GB \\
    \# HBM2 Stacks & 2 & 5 \\
    Est. \$-cost & \$263 & \$866 \\ \hline
    \end{tabular}
    \\
    \footnotesize{Note: We rename NVIDIA's L2 to LLC to compare with \ourarch{}.}
\end{table}

We now compare the \ourarchbase{} implementation, against an NVIDIA A100 GPU. We choose a relatively small batch size of 16 to contain the need for the high memory bandwidth and dollar cost of large batch training and also to stress \ourarch{} capability of extracting out parallelism when batch-level parallelism is low\footnote{At batch size 16, A100 utilization is less than 1\% for RNN-T, skewing our results, so we choose batch size 512 for RNN-T. Also, RNN-T's recurrent architecture lends to a small model size which afford very large batch execution without exorbitant memory and bandwidth needs.}. Table~\ref{tab:spec-comparison} shows a spec comparison of \ourarchbase{} to the A100. We build a simple cost model based on wafer/die costs published here (\$238 for 600$mm^2$ die)~\cite{die-cost-model} and publicly available information on GDDR6 chip cost (roughly \$90 for 8GB)~\cite{gddr6-cost}, and assume optimistically for A100 (that HBM costs the same). This indeed ignores cost of interposer, packaging etc. Figures~\ref{fig:perf-compare} shows the comparison of performance. \textbf{Overall, \ourarchbase{} achieves \gminferlarge{} / \gmtrainlarge{} geo-mean speed-up at batch 16 size inference / training with \powinferlarge{} / \powtrainlarge{} power efficiency.} 
We intentionally picked a design point that was iso-performance to the A100. \textbf{This makes clear the speedup we see with \ourarchbase{} comes from improving the utilization of compute resources by the same factor, with almost an order of magnitude lower energy.} \textbf{Energy efficiency of U2048 is much higher, because we rely on power-efficient microarchitecture targeted to DL exploiting reuse, data movement and orchestration, vs. a GPU that must adhere to a throughput-optimized execution model that constantly moves values in and out of main memory.}

To highlight the scope of \ourarch{}'s design space and flexibility, a chip optimized for transformer training would result in having 4096 cores, 512-bit SIMD, operating at 2.4GHz. This ViTr configuration is able to improve the energy efficiency gain over A100 from 6.3X to 8X for BERT training.

\subsection{Operator Analysis of MLPerf App Performance}\label{sec:shape-analysis}

\begin{table}
    \centering
    \caption{Analysis of Top 3-5 layers for each app. by op count}
    \label{tab:shape-table}
\small

\begin{tabular}{llrrrl}
\hline
\multicolumn{1}{c}{\textbf{Op}} & +TP & -TP & \multicolumn{1}{c}{\textbf{A100}} & \multicolumn{1}{c}{\textbf{Comments}} \\ \hline
\multicolumn{5}{c}{\textbf{Resnet50 (All 2D Conv.)}} \\ \hline
I 14$^2$ 256-\textgreater{}256 f=3 s1 & 82 & 68 & 14 & \tag{SS} \tag{LC} \\ 
I 56$^2$ 64-\textgreater{}64 f=3 s1 & 90 & 78 & 12 & \tag{LS} \tag{SC} \\ 
I 28$^2$ 128-\textgreater{}128 f=3 s1 & 80 & 82 & 16 & \tag{SS} \\ 
I 14$^2$ 256-\textgreater{}1024 f=1 s1 & 69 & 93 & 8 & \tag{F1} \tag{LC} \tag{A-} \\ 
I 14$^2$ 1024-\textgreater{}256 f=1 s1 & 66 & 72 & 9 & \tag{SS} \tag{BP} \\ 
T 14$^2$ 256-\textgreater{}256 f=3 s1 & 25 & 5 & 15 & \tag{DW} \tag{SS} \\ 
T 56$^2$ 64-\textgreater{}64 f=3 s1 & 61 & 6 & 11 & \tag{DW} \tag{LS} \tag{SC} \\ 
T 28$^2$ 128-\textgreater{}128 f=3 s1 & 65 & 6 & 14 & \tag{DW} \tag{LS} \tag{SC} \\ 
T 14$^2$ 256-\textgreater{}1024 f=1 s1 & 34 & 15 & 8 & \tag{DW} \tag{F1} \tag{LC} \\ 
T 14$^2$ 1024-\textgreater{}256 f=1 s1 & 67 & 16 & 9 & \tag{DW} \tag{F1} \tag{LC} \\ 
\hline
\multicolumn{5}{c}{\textbf{SSD-Resnet34 (All 2D Conv.)}} \\ \hline
I 150$^2$ 256-\textgreater{}256 f=3 s1 & 94 & 46 & 53 & \tag{LS} \tag{LC} \tag{GP} \\ 
I 150$^2$ 128-\textgreater{}128 f=3 s1 & 97 & 82 & 37 & \tag{LS} \tag{LC} \\ 
I 300$^2$ 64-\textgreater{}64 f=3 s1 & 99 & 74 & 26 & \tag{LS} \\ 
I 150$^2$ 128-\textgreater{}256 f=3 s1 & 98 & 67 & 43 & \tag{LS} \tag{LC} \\ 
I 150$^2$ 256-\textgreater{}512 f=3 s2 & 77 & 17 & 45 & \tag{S2} \\ 
T 38$^2$ 256-\textgreater{}256 f=3 s1 & 53 & 5 & 37 & \tag{DW} \tag{SS} \tag{LC} \\ 
T 38$^2$ 128-\textgreater{}128 f=3 s1 & 62 & 6 & 20 & \tag{DW} \tag{SS} \tag{LC} \\ 
T 75$^2$ 64-\textgreater{}64 f=3 s1 & 69 & 6 & 14 & \tag{DW} \tag{SS} \tag{LC} \\ 
T 38$^2$ 128-\textgreater{}256 f=3 s1 & 45 & 6 & 27 & \tag{DW} \tag{SS} \tag{LC} \\ 
T 38$^2$ 256-\textgreater{}512 f=3 s2 & 25 & 3 & 27 & \tag{DW} \tag{SS} \tag{LC} \\ 
\hline
\multicolumn{5}{c}{\textbf{UNet (All 3D Conv.)}} \\ \hline
I 32$^3$ 32-\textgreater{}32 f=3 s1 & 24 & 49 & 22 & \tag{BP} \tag{LS} \tag{SC} \\ 
I 32$^3$ 64-\textgreater{}32 f=3 s1 & 24 & 99 & 15 & \tag{BP} \tag{LS} \tag{SC} \\ 
I 32$^3$ 64-\textgreater{}64 f=3 s1 & 49 & 99 & 28 & \tag{BP} \tag{LS} \tag{SC} \\ 
I 32$^3$ 128-\textgreater{}64 f=3 s1 & 49 & 99 & 35 & \tag{BP} \tag{LS} \tag{LC} \\ 
I 32$^3$ 128-\textgreater{}128 f=3 s1 & 99 & 99 & 53 & \tag{BP} \tag{LS} \tag{LC} \\ 
T 32$^3$ 32-\textgreater{}32 f=3 s1 & 27 & 8 & 16 & \tag{DW} \tag{LS} \tag{SC} \\ 
T 32$^3$ 64-\textgreater{}32 f=3 s1 & 39 & 8 & 19 & \tag{DW} \tag{LS} \tag{SC} \\ 
T 32$^3$ 64-\textgreater{}64 f=3 s1 & 57 & 8 & 35 & \tag{DW} \tag{LS} \tag{SC} \\ 
T 32$^3$ 128-\textgreater{}64 f=3 s1 & 71 & 8 & 43 & \tag{DW} \tag{LS} \tag{LC} \\ 
T 32$^3$ 128-\textgreater{}128 f=3 s1 & 94 & 8 & 57 & \tag{DW} \tag{LS} \tag{LC} \\ 
\hline
\multicolumn{5}{c}{\textbf{BERT-Large}} \\ \hline
I Fc(2848x1024x1024) & 87 & 76 & 32 & \tag{LM} \tag{LN} \tag{LK} \\ 
I Fc(2848x4096x1024) & 91 & 31 & 93 & \tag{LM} \tag{LN} \tag{LK} \\ 
I Fc(2848x1024x4096) & 90 & 52 & 31 & \tag{LM} \tag{LN} \tag{LK} \\ 
I Mm(178x178x64) & 27 & 27 & 8 & \tag{LL} \tag{SM} \tag{SN} \\ 
I Mm(178x64x178) & 14 & 14 & 6 & \tag{LL} \tag{SM} \tag{SN} \\ 
T Fc(4064x1024x1024) & 83 & 14 & 95 & \tag{TP} \tag{LM} \tag{LN} \tag{LK} \\ 
T Fc(4064x4096x1024) & 83 & 14 & 47 & \tag{TP} \tag{LM} \tag{LN} \tag{LK} \\ 
T Fc(4064x1024x4096) & 83 & 14 & 93 & \tag{TP} \tag{LM} \tag{LN} \tag{LK} \\ 
T Mm(254x254x64) & 2 & 2 & 9 & \tag{TP} \tag{LL} \tag{SM} \tag{SN} \\ 
T Mm(254x64x254) & 5 & 5 & 17 & \tag{TP} \tag{LL} \tag{SM} \tag{SN} \\ 
\hline
\multicolumn{5}{c}{\textbf{RNN-T}} \\ \hline
I Lstm(512x4096x2048) & 65 & 18 & 22 & \tag{EW} \tag{LN} \tag{LK} \\ 
I Lstm(512x4096x3072) & 50 & 36 & 24 & \tag{EW} \tag{LN} \tag{LK} \\ 
I Lstm(512x4096x1264) & 91 & 56 & 16 & \tag{EW} \tag{LN} \tag{LK} \\ 
I Lstm(512x1280x640) & 48 & 65 & 6 & \tag{EW} \tag{SK} \\ 
I Fc(512x1344x512) & 46 & 71 & 11 & \tag{EW} \tag{BP} \\ 
T Lstm(512x4096x2048) & 41 & 11 & 19 & \tag{TP} \tag{EW} \tag{LN} \tag{LK} \\ 
T Lstm(512x4096x3072) & 41 & 11 & 23 & \tag{TP} \tag{EW} \tag{LN} \tag{LK} \\ 
T Lstm(512x4096x1264) & 36 & 11 & 16 & \tag{TP} \tag{EW} \tag{LN} \tag{LK} \\ 
T Lstm(512x1280x640) & 47 & 14 & 5 & \tag{TP} \tag{EW} \tag{SK} \\ 
T Fc(512x1344x512) & 42 & 14 & 7 & \tag{TP} \tag{EW} \\ 
\hline
\end{tabular}

\end{table}

We now dive deeper into each application by conducting a layer-wise analysis of MLPerf and extract out and analyze important, architecture-agnostic behaviors. We also discuss how \ourarch{} is able to achieve high performance given these behaviors. Table~\ref{tab:shape-table} summarizes our findings for the top 3-5 operator shapes in each network by percentage of total op count. Each row is one (I) inference or (T) training layer of the network with utilization as a percentage of peak compute throughput for \ourarchbase{} with the Transpose Engine enabled (+TP), with the Transpose Engine disabled (-TP) and for the A100. The symbols we use have the following meaning. \tag{LC}/\tag{SC}, \tag{LS}/\tag{SS}: Large/Small Channel, Spatial Conv.
\tag{F1}: Filter size = 1 Conv. \tag{DW}: Conv. backprop for weights. \tag{BP}: Bad placement caused by tiling. \tag{LM}/\tag{SM}, \tag{LN}/\tag{SN}, \tag{LK}/\tag{SK}: Large/Small M, N, K matmul. \tag{TP}: Transposed matmuls for backprop. \tag{EW}: Elementwise operations. \textbf{We identify the fundamental application behaviors that are key to achieving high performance and efficiency. The identification and explanation of these behaviors are a contribution that is architecture agnostic and to our knowledge, the most comprehensive such analysis.}

\myparagraph{1. Large Channel Convolution} convolutions have a large channel \tag{LC} dimensions, making inner matrix multiplies amenable to both \ourarch{}'s SIMD unit and A100's TensorCore. Often channel dimensions are large enough that \ourarch{}'s transpose engine becomes unnecessary for performance. When spatial dimensions are small \tag{SS}, additional parallelism is extracted from output channels, putting more pressure on the transpose engine. In this case, we also observe that the A100 suffers in utilization; likely due to the fact the inner tile dimensions end up not filling the relatively large TensorCore.

\myparagraph{2. Large Spatial Convolution} convolutions have high parallelism from splitting work on spatial dimensions \tag{LS}, as exemplified by UNET's large spatial convolutions. We can also see the A100 appears to require both large spatial and channel dimensions to extract high utilization -- both L11 and L21 have a large amount of spatial parallelism available, but L21 has much smaller channel count, which is likely why A100's performance suffers.

\myparagraph{3. Unit-Filter Convolution} convolutions with one filter pixel \tag{F1} degenerate into a large matrix multiply (M = spatial dimension, N = output channel, K = input channel). In this case, \ourarch{} is able to perform quite well even without the transpose engine since typically channel dimensions are large. When M >> N, we find that communication becomes the bottleneck since arithmetic intensity drops.

\myparagraph{4. Convolution Backpropagation} Backpropagation for convolution is difficult for two reasons: 1) matrix multiply layouts are transposed, altering minimum tiling requirements, and 2) backpropagation for weights \tag{DW} cannot parallelize on spatial dimensions. With this parallelism gone, \ourarch{} relies on the channel dimensions, and when this is insufficient, partial gradients are computed over the batch dimension and summed together with in-LLC reduction as discussed in Section~\ref{sec:mapping}. It is likely that A100 suffers from a similar problem. 

\myparagraph{5. Tiling Effects on Placement} In some cases \tag{BP}, we observe a decrease in efficiency when using the transpose engine in \ourarch{}. In all cases we analyzed, this was not due to under-utilization of the SIMD engine, but instead the change in tiling factors as a result of using the transpose engine caused work placement to produce worse communication patterns. This is not a problem for \ourarch{} since we can either naively just switch off the transpose engine, or fine-tune the placement by adjusting tiling factors.

\myparagraph{6. Large Matrix Multiplies} The top 3 layers for BERT are matrix multiplies with large M \tag{LM}, large N \tag{LN}, large K \tag{LK}. These kinds of shapes are easy for both SIMD and TensorCore, so it's not surprising that both A100 and \ourarch{} perform well. It does appear the A100 requires even larger dimensions than L31 (for example) to achieve peak throughput (L32, L33). Additionally, similar to convolution, backward passes for matrix multiply are also matmuls but with transposed inputs \tag{TP}.

\myparagraph{7. Odd-shaped Batch Matrix Multiplies} BERT's self attention layers employ batch matrix multiplies with large outer L dimension \tag{LL} and relatively small M \tag{SM} and small N \tag{SN} dimensions. For these matrix multiplications, 1) the outer batch dimension afford embarrassing parallelism at the expense of interconnect pressure due to drop in arithmetic intensity 2) small and irregular M and N dimensions make it difficult to extract further parallelism meaning \ourarch{}'s transpose engine has little effect and 3) layout for this matrix multiply cannot be tuned for inference since neither input is a stored weight.

\myparagraph{8. LSTM with low parallelism} The LSTM layer can be thought of as a linear layer over both the input and recurrent state, followed by a relatively complex element-wise operation. \ourarch{} leverages previously discussed techniques for computing the matrix multiples in the linear layers efficiently. \ourarch{} additionally employs careful in-memory layout of tensors to allow for in-LLC elementwise \tag{EW} operation -- meaning the LSTM gates are all computed over local data in a tile's LLC slice, requiring no data movememnt after the linear layers.

\begin{table}[]
    \centering
    \caption{Analysis of related academic DL accelerators on basis of their efficiency when observed behaviors are present.}
    \label{tab:behav-qual-comp}
\begin{tabular}{ccccc}
Behav.                 & Simba  & EyerissV2 & MAGNet & SIGMA  \\ \hline
\multicolumn{1}{c|}{1,2,3} & Med-Hi     & Med-Hi        & Hi     & Hi     \\
\multicolumn{1}{c|}{4} & Unsup. & Unsup.    & Unsup. & Lo     \\
\multicolumn{1}{c|}{5} & Var.   & Var.      & Var.   & Var.   \\
\multicolumn{1}{c|}{6} & Hi     & Hi        & Hi     & Hi     \\
\multicolumn{1}{c|}{7} & Lo     & Hi        & Lo     & Med-Hi \\
\multicolumn{1}{c|}{8} & Med     & Unsup.    & Unsup. & Med    \\ \hline
\end{tabular}
\footnotesize{Behavior \#5 depends on shape, the architecture, and mapping strategy and so is difficult to say how tiling factors impact each design.}
\end{table}

\subsection{Qualitative Analysis of Academic Architectures}
To conclude this study, we present a qualitative analysis of academic DL accelerator designs on whether they would perform well for each of these behaviors. Table~\ref{tab:behav-qual-comp} summarizes our findings, which we now break down for each of the four architectures we study. Note that these still suffer from the unaddressed compilability challenge.

Simba only evaluates linear layers but should be able to perform convolutions. Given a vector width of 8, it should perform most convs. in MLPerf well but without additional data orchestration features, will suffer for some shapes. Being inference-only, training convolution is unsupported. For similar reasons, Simba is likely to perform well for large matmuls, but suffer for the irregular batch matmuls seen in transformers. Simba is likely to perform just okay for LSTMs since it will rely exclusively on batch parallelism, and only its Global PE is able to do near-memory reduction operations. Overall, Simba's dynamic NoC and multicast capability from some units make it amenable to easy data movement, in addition, work placement, but its fixed width SIMD units and lack of support for training cause it to fall short on coverage and performance of MLPerf.

EyerissV2 is also an inference engine focusing on convolution. It is likely to support convolutions in MLPerf quite well with its row-stationary dataflow, but does not support training. Since matmuls are a degenerate case of convolution, EyerissV2 should also support matmul layers, likely working well for large matmuls. EyerissV2 would also likely work well for the irregular matmuls in transformers since its PEs are very fine-grained, making data-orchestration much easier at the PE level. EyerissV2 does not support activation functions, though so would not be able to run LSTM ops. EyerissV2's interconnect is statically programmed by software, and if the routing needs of an application exceed hardware resources, software will be unable to route data. Overall, EyerissV2 has good data orchestration at the expense of complex, software controlled data movement. It has no support beyond DNN and CNN inference, so falls short on DL coverage.

MAGNet is an RTL generator which intentionally echews coverage in order to attain the highest possible performance and efficiency for a single application. It employs conventional SIMD execution, meaning it suffers from the data orchestration problems as Simba, but should support nice convolutions and matmuls well, but will likely suffer for irregular batch matmul shapes. It employs a dynamic interconnect so data movement and work placement is a relatively easy lift for software, but without multicast or the ability to ``push'' data, though, it has to route requests as well as data. Overall, MAGNet was not designed for coverage, and lacks data orchestration and movement techniques to achieve high performance for irregular shapes, as well as the ability to perform training.

SIGMA's FlexDPE is capable of very flexible data orchestration at the core level. The tradeoff is a relatively complex fan-out/in network to deliver elements to hardware execution units, which we find to be over-engineered for DL applications (as exemplified by MLPerf). It has a NoC similar to MAGNet. It will likely perform well for nice matrix multiplies and convolutions as a result. For conv. backprop., it will likely suffer since it cannot perform near-data reductions to reduce communication pressure. It's performance for irregular matmuls will depend on whether the dimensions can fill the relatively large FlexDPE size. SIGMA will likely support LSTM operations about as well as Simba or A100. Overall, SIGMA has the highest coverage of the accelerators we study, but lacks architectural features for optimizing data movement, exacerbating its compilability limitation.

\subsection{\ourarch{}'s Roadmap to the next 100X}\label{sec:sens-scale}

\begin{figure}
    \centering
    \includegraphics[width=1.0\columnwidth]{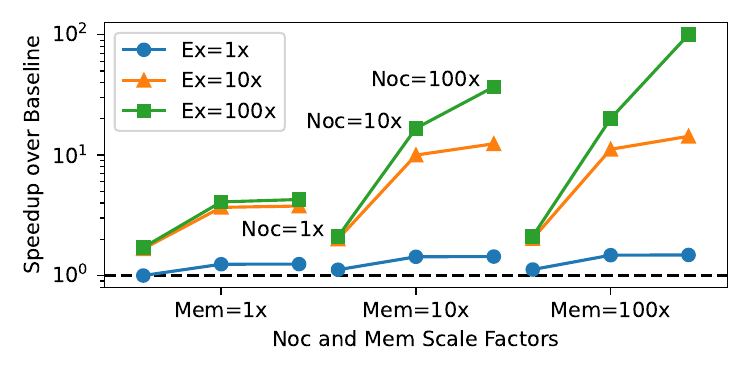}
    \caption{Overall performance sensitivity to Memory, Communication network, and Compute engine.}
    \label{fig:sens}
\end{figure}

To examine the scalability of \ourarch{}, we simulate the performance improvement from scaling each of compute, communication, and memory bandwidth by a factor of 1X, 10X, and 100X, separately and together (a total of 27 design points). Figure~\ref{fig:sens} shows these speedups normalized to  \ourarchbase{}. The following insights emerge. 
 \textbf{i) With additional memory bandwidth alone, at best 25\% speedup is possible}.
\textbf{ii) Conversely, since utilization is already high, improving memory bandwidth and NoC alone or together provides limited speedups.}
 \textbf{iii) Surprisingly, 10X in bandwidth and compute, with no change to the NoC, provides about 9X speedups, meaning the push-based NoC architecture scales.} Emerging packaging solutions could make this direction realistic to achieve. \textbf{iv) Getting speedups beyond 10X also seems surprisingly possible and not limited by application
inherent characteristics.} Microarchitecture/architecture techniques that create an effective increase of 100X in memory bandwidth (main memory caches), fast NoCs (photonic like Corona~\cite{vantrease2008corona}) could help realize these design points in a practical way.




\section{Conclusion}
This paper identifies the four key problems which must be solved for a new DL accelerator to be successful. Through a fresh perspective of extending established multicore SIMD architecture, we develop \ourarch{}, which solves these four problems in a balanced way, lending to high performance and coverage of modern DL applications. We provide
a surprisingly effective approach that outperforms GPUs by large integer factors,
with a substantially lower silicon footprint design. The analysis and key behaviors we identified are leveraged by \ourarch{} to achieve this, and have implications for future architectures as well.



\bibliographystyle{plain}
\bibliography{references}


\end{document}